\documentclass{desyproc}

\def \F{\mathcal{F}}

\begin{document}
\title{Axion Search with Ring Cavity Experiment}

\author{{\slshape Ippei Obata$^1$, Tomohiro Fujita$^2$, Yuta Michimura$^3$}\\[1ex]
$^1$Institute for Cosmic Ray Research, University of Tokyo, Kashiwa 277-8582, Japan\\
$^2$Department of Physics, Kyoto University, Kyoto, 606-8502, Japan\\
$^3$Department of Physics, University of Tokyo, Bunkyo, Tokyo 113-0033, Japan}
\contribID{Lindner\_Axel}

\confID{20012}  
\desyproc{DESY-PROC-2018-03}
\acronym{Patras 2018} 
\doi  

\maketitle

\begin{abstract}
We suggest a novel experimental method to search for axion dark matter with an optical ring cavity.
 Our cavity measures the difference of the resonant frequencies between two circular-polarizations of the laser beam.
 Its technical design adopts double-pass configuration to realize a null experiment and reject environmental common-mode noises. 
 We reveal that it can probe the axion-photon coupling constant with a broad range of axion mass $10^{-17} \text{eV} \lesssim m \lesssim 10^{-10} \text{eV}$, up to several orders of magnitude beyond the current limits.
 We expect that this cavity experiment establishes a new window to develop the axion research.
  \end{abstract}

\section{Introduction}

We propose a new experiment to search for the coupling of photon to axion dark matter \cite{Obata:2018vvr}.
The axion, motivated by higher dimensional theories, is known to be one of the best candidate of dark matter \cite{Marsh:2015xka}.
Its coherent mode oscillating around the minimum of its potential provides a small difference in the phase velocity between the left-handed and the right-handed photon.
 Inspired by this photon birefringence effect, we suggest that the optical ring cavity is useful to detect such a small deviation of the phase velocity.
 Ring cavity experiments have been recently emerged to test the parity-odd Lorentz violation in the photon sector~\cite{Baynes:2012zz}.
They have measured the variation of the resonant frequency depending on the direction of the light path. 
Similar technique can be applied for our purpose, 
because the resonant frequency of the cavity shifts depending on the polarization of photons, provided that the dark matter axion is coupled to photon.
The dark matter axion predicts the phase velocities of the left and right-handed polarized photon shift with the opposite signs and the same magnitude.
Therefore such shifts of the resonant frequencies of the polarized laser
in the optical cavity are the measurement target in our experiment.
The sensitivity curve is in principle determined only by quantum shot noise by virtue of the double-pass configuration and hence we can achieve the great sensitivity level for the detection of the axion-photon coupling constant.
We have demonstrated that it can reach sensitivities beyond the current constraints by several orders of magnitude.

\section{Phase velocities of photons}

In this section, we estimate the phase velocities of two circular-polarized photons coupled with the axion dark matter.
The axion-photon coupling term is written as $g_{a\gamma}aF_{\mu\nu}\tilde{F}^{\mu\nu}/4$, where $a(t)$ is the axion field value,  $F_{\mu\nu} \equiv \partial_\mu A_\nu - \partial_\nu A_\mu$ is the field strength of vector potential $A_\mu$, and $\tilde{F}_{\mu\nu} \equiv \epsilon_{\mu\nu\rho\sigma}\partial_\rho A_\sigma/2$ is its dual.
Here, we adopt the temporal gauge $A_0 = 0$ and the Coulomb gauge $\nabla\cdot \bf{A} = 0$.
 Then the equation of motion (EoM) for gauge field reads
\begin{equation}
\ddot{A}_i - \nabla^2A_i + g_{a\gamma}\dot{a}\epsilon_{ijk}\partial_j A_k = 0 \ , \quad ``\cdot" = d/dt \, .
\end{equation}
The present background axion field is given by $a(t) = a_0\cos(m t + \delta_\tau(t))$ with its constant amplitude $a_0$, its mass $m$ and a phase factor $\delta_\tau(t)$.
Note that we assume $\delta_{\tau}$ to be a constant value within the coherence timescale of dark matter $\tau = 2\pi/(mv^2) \sim 1\,\text{year}(10^{-16}\text{eV}/m)$.

Decomposing $A_i$ into two helicity modes with the wave number $\bf{k}$, we find EoMs for the two polarization modes $A^\pm_k$ as
\begin{equation}
\ddot{A}^\pm_k + \omega_\pm^2A^\pm_k = 0\,, \quad \omega_\pm^2 \equiv k^2\left( 1 \pm \dfrac{g_{a\gamma}a_0 m}{k}\sin(m t + \delta_\tau) \right)
\end{equation}
and obtain the difference of their phase velocities $\delta c \equiv |c_+ - c_-|$.
Since the coupling constant $g_{a\gamma}$ is tiny, $\delta c$ is approximately given by
\begin{equation}
\delta c \simeq \dfrac{g_{a\gamma}a_0 m}{k}\sin(m t + \delta_\tau) \equiv \delta c_0\sin(m t + \delta_\tau)\,.
\end{equation}
 Therefore, with the wavelength $\lambda = 2\pi/k = 1550~\text{nm}$ we can estimate
\begin{equation}
\delta c_0
\simeq 
3\times 10^{-24}\left(\dfrac{g_{a\gamma}}{10^{-12}~\text{GeV}^{-1}}\right)\,,
\end{equation}
where we used the present energy density of the axion dark matter, 
$\rho_a = m^2a_0^2/2 \simeq 0.3~\text{GeV}/\text{cm}^{3}$.


\section{Optical ring cavity search for axion dark matter}

In this section, we describe our experiment to detect $\delta c$ caused by the axion dark matter and calculate the sensitivity to the axion-photon coupling constant.
The setup of our experiment is schematically illustrated in Figure 1.
First, we create a incident laser beam which is circularly polarized by a 1/4 waveplate. 
The incident beam to the cavity is partially reflected
by the input mirror and goes to the photodetector A, while the other
part enters the cavity which eventually goes to either the photodetector A or the mirror on the far right. 
And the beam which is reflected from the mirror on the far right 
is partially reflected into the photodetector B
or re-enters the cavity, and finally some part of the beam goes into the photodetector B.
Note that the beam changes its polarization each time when it is reflected by a mirror.
Considering this effect, we stretch the bow-tie optical path in the longitudinal direction in order to make the dominant polarization states at each photodetector.
For instance, we can make the laser beam entering the detector A right-handed polarization most of the time, while the other laser beam left-handed polarization going into the detector B.
Without the phase velocity modulation $\delta c$ given by the axion dark matter,
the resonant frequency would not depend on the circular-polarizations. 
Therefore our setup works as a null-experiment sensitive to the axion-photon
coupling.
%
\begin{figure}[tbp]
\begin{center}
\begin{tabular}{c}
\begin{minipage}{0.5\hsize}
\begin{center}
\includegraphics[width=0.95\textwidth]{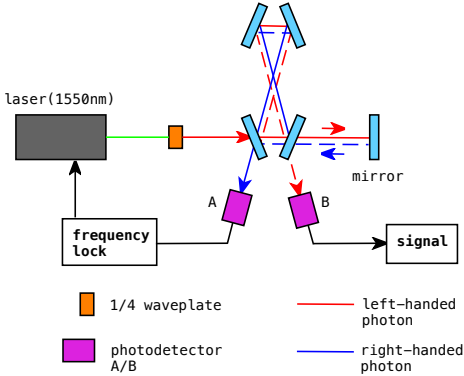}
\end{center}
\end{minipage}
\label{fig: ringcavity}
\begin{minipage}{0.5\hsize}
\begin{center}
\caption{The layout of our double-pass bow-tie cavity.
 The left-handed beam (solid line) is injected to the resonant cavity,
 while the transmit beam reflected by the mirror on the far right goes to the cavity as the right-handed beam (dashed line).
 The photodetector A is used to lock the laser frequency at the resonant  frequency for the injected beam from the left, and the photodetector B monitors
 the modulation of the resonant frequency difference of two optical paths from the beam coming from the right.}.
\end{center}
 \end{minipage}
\end{tabular}
\end{center}
\end{figure}
%

 In our setup, most of the environmental noises are also cancelled due to the
double-pass configuration~\cite{Cusack:2002dw}, because the second
error signal observes only the difference in the resonant frequency between the two counter-propagating optical paths in the cavity and their common fluctuations become irrelevant.
 Then the primary source of noise is the quantum shot noise.
One-sided spectrum of the shot noise of an optical ring cavity is written as \cite{Kimble:2000gu}
\begin{equation}
  \sqrt{S_{\rm{shot}}} = \sqrt{\frac{\lambda}{4 \pi P} \left( \frac{1}{t_r^2} + \omega^2 \right)} \label{eq: shotN} \ , \qquad t_r = \frac{L \F}{\pi} \ ,
\end{equation}
where $\lambda$ is the laser wavelength, $P$ is the input power, and $\omega$ is the angular frequency which is the axion mass $m$ in our case.
 Note that the quantum radiation pressure noise is cancelled out by our double-pass configuration.
 The averaged round-trip time $t_r$ is characterized by the cavity round-trip length $L$ and the finesse $\F$.
 If our measurement is limited by the shot noise, the signal-to-noise ratio (SNR) improves with the measurement time $T^{1/2}$ as long as the axion oscillation is coherent for $T \lesssim \tau$.
 When the measurement time becomes longer than this coherence time $T > \tau$, the phase $\delta_\tau$ is not constant any more and $\delta_\tau$ will behave as a random variable staying constant for each period of $\tau$.
As a consequence, the growth of the SNR with the measurement time changes as
$(T\tau)^{1/4}$ \cite{Budker:2013hfa}.
Therefore the sensitivity to $\delta c_0 /c$ is limited by
\begin{align}
\frac{\delta c_0}{c}
\lesssim
\begin{cases}
\dfrac{2}{\sqrt{T}} \sqrt{S_{\rm{shot}}} & (T \lesssim \tau) \\
\dfrac{2}{(T\tau)^{1/4}} \sqrt{S_{\rm{shot}}} & (T \gtrsim \tau) \\
\end{cases} \quad \longleftrightarrow \quad  g_{a \gamma } &\lesssim 
 \begin{cases}
10^{12} \sqrt{\dfrac{S_{\rm{shot}}}{T}}~[1/\rm{GeV}] & (T \lesssim \tau) \\
10^{12}\sqrt{\dfrac{S_{\rm{shot}}}{(T\tau)^{1/2}}}~[1/\rm{GeV}] & (T\gtrsim \tau) \\
\end{cases}\,.
\end{align}

In this experiment, we search for the axion dark matter with the mass $m\lesssim 10^{-10}$eV
corresponding to the frequency range $f \simeq 2.4\, {\text Hz} (m/10^{-14}\text{eV})$.
Figure 2 shows the sensitivity of our experiment to the axion-photon coupling constant for different configurations.
Here we set $\lambda=1550$~nm and assumed $T= 1~{\rm year} = 3 \times 10^{7}~{\rm sec}$.
With feasible parameters we can achieve a sensitivity level $g_{a\gamma} \simeq 3 \times 10^{-13} ~\text{GeV}^{-1}$ for $m \lesssim 10^{-16} ~\text{eV}$, which is below the current constraints from axion helioscope experiments, SN1987A and {\sl Chandra} X-ray observations.
Moreover, with more optimistic parameters, our cavity can reach $g_{a\gamma} \simeq 3 \times 10^{-16} ~\text{GeV}^{-1}$ for $m \lesssim 10^{-16} ~\text{eV}$ which will be the best sensitivity among the proposed axion search experiments in this mass range.
 We note here that various technical noises at low frequency should be further investigated to determine the sensitivity for lower mass range of the axion.
 We leave this issue for future work.
\label{fig: bound}
\begin{figure}[tbp]
\begin{center}
\begin{tabular}{c}
\begin{minipage}{0.5\hsize}
\begin{center}
\includegraphics[width=0.85\textwidth]{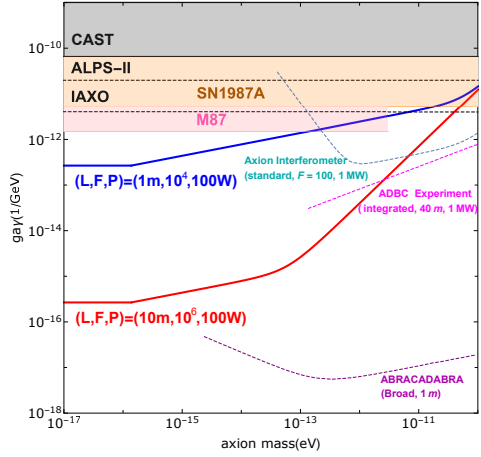}
\end{center}
\end{minipage}

\begin{minipage}{0.5\hsize}
\begin{center}
\caption{The sensitivity curves for the axion-photon coupling constant $g_{a\gamma}$ with respect to the axion mass $m$.
 The solid blue (red) line shows the sensitivity of our experiment $(L, F, P) = (1(10)~\text{m}, 10^4(10^6), 10^2(10^2)~\text{W})$.
 The gray band represents the current limit from CAST~\cite{Zioutas:2004hi}.
 The dashed black lines are the prospected limits of IAXO \cite{Vogel:2013bta} and ALPS-II \cite{Ehret:2009sq} missions.
 The dashed blue, magenta and purple lines show the proposed reaches of axion optical interferometers \cite{DeRocco:2018jwe}, birefringent cavities \cite{Liu:2018icu} and ABRACADABRA magnetometer \cite{Kahn:2016aff}.
 The orange and pink bands denote the astrophysical constraints from the cosmic ray observations of SN1987A \cite{Payez:2014xsa} and radio galaxy M87 \cite{Marsh:2017yvc}.
}
\end{center}
 \end{minipage}
\end{tabular}
\end{center}
\end{figure}
%

\section{Acknowledgments}
\label{Acknowledgement}

 In this work, YM and TF are supported by the JSPS Grant-in-Aid for Scientific Research (B) No.~18H01224 and Grant-in-Aid for JSPS Research Fellow No.~17J09103, respectively.


\begin{footnotesize}
\begin{footnotesize}

\end{footnotesize}

\end{footnotesize}


\end{document}